\documentclass[]{pasj01}
\draft

\begin{document} 
\Received{}
\Accepted{}

\title{Suzaku observations of two diffuse hard X-ray source regions, G22.0$+$0.0 and G23.5$+$0.1}

\author{
Shigeo \textsc{Yamauchi}\altaffilmark{1,$\ast$},
Mayu \textsc{Sumita}\altaffilmark{1},
and
Aya \textsc{Bamba}\altaffilmark{2}
}
 \altaffiltext{1}{Department of Physics, Nara Women's University, Kitauoyanishimachi, Nara 630-8506}
\email{yamauchi@cc.nara-wu.ac.jp}
 \altaffiltext{2}{Department of Physics and Mathematics, Aoyama Gakuin University, \\
5-10-1 Fuchinobe, Chuo-ku, Sagamihara, Kanagawa 252-5258}


\maketitle

\begin{abstract}
G22.0$+$0.0 and G23.5$+$0.1 are diffuse hard X-ray sources discovered in the ASCA Galactic Plane Survey.
We present Suzaku results of spectral analysis for these sources.
G22.0$+$0.0 is confirmed to be a largely extended emission. 
The spectra were represented by a highly absorbed power-law model 
with a photon index of 1.7$\pm$0.3 and a moderately absorbed thermal emission with a temperature of 0.34$^{+0.11}_{-0.08}$ keV.
The difference in the $N_{\rm H}$ values between the two components 
suggests that the thermal component is unrelated with the power-law component and 
is a foreground emission located in the same line-of-sight.
G23.5$+$0.1 is an extended source with a size of $\sim$\timeform{3'.5}.
The spectra were fitted with an absorbed power-law model with a photon index of 2.4$^{+0.5}_{-0.4}$.
The spatial and spectral properties show that both are candidates of old pulsar wind nebulae (PWNe).
In addition to the extended sources, we analyzed spectra of three point sources found in the observed fields.
Based on the spectral features, we discuss origin of the sources.
\end{abstract}

\section{Introduction}

Due to the large extinction, foreground radiation, and the contamination of nearby bright sources,
it is difficult to detect faint sources located on the Galactic plane.
Therefore, sample or information of the faint objects inside the Galactic plane has been limited. 
In the X-ray band above 3 keV, the interstellar medium is essentially transparent through the Galactic plane. 
Thus, observations in the hard X-ray band will be able to discover new faint X-ray sources on the Galactic plane. 
The ASCA satellite was the first satellite having an imaging capability up to 10 keV with an angular resolution of $\sim$3$'$,
expressed in terms of the half power diameter (HPD) \citep{Tanaka1994}. 
We carried out ASCA Galactic Plane Survey (AGPS) \citep{Yamauchi2002}
and detected many X-ray sources including unidentified sources \citep{Sugizaki2001,Sakano2002}.

G22.0$+$0.0 and G23.5$+$0.1 are hard X-ray sources discovered in the AGPS \citep{Sugizaki2001,Ueno2005,Ueno2006}.
The X-ray images showed that G22.0$+$0.0 has an extended emission 
with a radius of $\sim$5$'$, while G23.5$+$0.1 has an arc-like structure \citep{Ueno2006}.
The spectra were represented by an absorbed power-law (PL) model 
with a photon index $\Gamma$=1.0$^{+0.7}_{-0.3}$ and $N_{\rm H}$$<$1.3$\times10^{22}$ cm$^{-2}$ for G22.0$+$0.0
and $\Gamma$=2.5$^{+1.1}_{-0.8}$ and $N_{\rm H}$=(6$^{+4}_{-3}$)$\times10^{22}$ cm$^{-2}$ 
for G23.5$+$0.1 (90\% confidence level) \citep{Ueno2005}. 
A non-thermal supernova remnant (SNR) scenario was proposed \citep{Ueno2005, Ueno2006}, but
their nature was unsolved.

XMM-Newton, with a better angular resolution (half energy width $\sim$17$''$ in orbit, 
XMM-Newton Users Handbook\footnote{http://xmm.esac.esa.int/external/xmm\_user\_support/documentation/uhb/index.html}), 
observed the G23.5$+$0.1 region in a snapshot survey of plerionic and composite SNRs 
in the Galactic plane region in 2006 \citep{Kargaltsev2012} 
and a follow up observation of SGR J1833$-$0832 in 2010 \citep{Esposito2011}.
The observations clearly found an extended source around 
the radio pulsar B1830$-$08=PSR J1833$-$0827 ($P$=85.3 ms, $\tau$={\it P}/2{\it \.{P}}=147 kyr, $d$=4.66 or 5.7 kpc, 
Hobbs et al. 2004, 2005; \cite{Kargaltsev2012}).
The extended source is proposed to be a pulsar wind nebula (PWN) generated by B1830$-$08 \citep{Esposito2011,Kargaltsev2012}.
The spectra of the whole source were fitted with an absorbed PL model with
$\Gamma$=2.3$\pm$0.8 and $N_{\rm H}$=(3.9$\pm$1.9)$\times10^{22}$ cm$^{-2}$ (68\% confidence level) \citep{Kargaltsev2012},
while the spectra of B1830$-$08 and the surrounding nebula were well represented by the absorbed PL model
with $\Gamma$ of 1.9$^{+0.7}_{-0.6}$ and 1.7$^{+0.5}_{-0.4}$ (68\% confidence level), respectively \citep{Esposito2011}.
 
Since photon statistics in the ASCA and XMM-Newton observations were limited, 
their spectral parameters have large uncertainties.
In order to reveal the spectral properties of the faint diffuse X-ray sources and to pin down their nature, 
the statistically good data are required. 
X-ray Imaging Spectrometers (XISs) onboard Suzaku has better spectral resolution and lower and more stable
detector background than the previous X-ray satellites \citep{Koyama2007,Mitsuda2007}.
Therefore, it is a suitable facility to study diffuse X-ray sources.  
We carried out follow up observations of G22.0$+$0.0 and G23.5$+$0.1 with Suzaku.
Throughout this paper, the quoted errors are at the 90\% confidence level.

\section{Observations and data reduction}

%
\begin{table*}[t]
\caption{Observation logs.}
\begin{center}
\begin{tabular}{lcccc} \hline  
Field & Observation ID & Pointing Position                   & Observation time (UT)& Exposure \\
            &                             & ($\alpha$, $\delta$)$_{\rm J2000.0}$ & Start -- End & (ks) \\
\hline 
G22.0$+$0.0& 505025010 & (\timeform{277.D8263}, \timeform{-9.D7167}) & 2010-04-16 14:27:26 -- 2010-04-17 17:27:12& 50.5\\
G23.5$+$0.1& 505026010 & (\timeform{278.D4828}, \timeform{-8.D3633}) & 2010-10-20 13:34:39 -- 2010-10-22 01:45:11& 49.0\\
\hline \\
\end{tabular}
\end{center}
\end{table*}

Suzaku observations of G22.0$+$0.0 and G23.5$+$0.1 were carried out 
with the XIS \citep{Koyama2007}
placed at the focal planes of the thin foil X-ray Telescopes (XRTs, \cite{Serlemitsos2007}).
The angular resolution (HPD) of the XRTs are \timeform{1.'8}--\timeform{2.'3} \citep{Serlemitsos2007}.
The XIS is composed of 4 sensors.
XIS 1 is a back-side illuminated (BI) CCD, while
the other three XIS sensors (XIS 0, 2, and 3) are front-side illuminated (FI) CCDs.
The field of view (FOV) of the XIS is \timeform{17.'8}$\times$\timeform{17.'8}.
Since the XIS 2 stopped working on 2006 November 9, 
we used the data obtained with XIS 0, XIS 1, and XIS 3.
A small fraction of the XIS 0 area was not used because of the data damage 
possibly due to an impact of micro-meteorite on 2009 June 23.
The XIS was operated in the normal clocking mode with a time resolution of 8 s.
The spectral resolution of the XIS was degraded due to the radiation of cosmic particles.
In order to restore the XIS performance, the spaced-row charge injection (SCI) technique was applied.
Details of the SCI technique are given in \citet{Nakajima2008} and \citet{Uchiyama2009}.

Data reduction and analysis were made with the HEAsoft version 6.16. 
The XIS pulse-height data for each X-ray event were converted to 
Pulse Invariant (PI) channels using the {\tt xispi} software 
and the calibration database version 2014-07-01.
We rejected the data acquired at the South Atlantic Anomaly, 
during the earth occultation, and at the low elevation angle 
from the earth rim of $<5^{\circ}$ (night earth) and $<20^{\circ}$ (day earth).  
After removing hot and flickering pixels,
we used the grade 0, 2, 3, 4, and 6 data.
The observation logs are listed in table 1.

\section{Analysis and results}

\subsection{Image}

Figure 1 shows X-ray images of the G22.0$+$0.0 and G23.5$+$0.1 fields in the 0.7--8 keV energy band.
For maximizing the photon statistics, the data of XIS 0, 1, and 3 were added.
In the G22.0$+$0.0 field, a largely extended emission with a diameter of $\sim$10$'$ (hereafter Src1) is found at the FOV center, while 
in the G23.5$+$0.1 field, several sources are seen; 
three bright point-like sources at the north (hereafter Src3), near to the south edge (hereafter Src4), and at the center (hereafter Src5)
and an extended source with a diameter of $\sim$\timeform{3'.5} at the southwest (hereafter Src2).
Comparing the observed radial profile with the point-spread-function (PSF) simulated using {\tt xissim}, 
we confirmed 
that Src1 and 2 have extents over the PSF and the others are likely to be point sources.

In addition, some faint sources are found in the images
(for example, near to the southwest corner in the G22.0$+$0.0 field and northeast of Src5).
Due to the limited photon statistics (the signal-to-noise ratio$<$10), we were not able to determine their spectral parameters well. 
Therefore, we do not treat them in this paper.

The sky positions of Src1--5 are listed in table 2. 
For Src1 and Src2, we defined the brightest position in the Suzaku image as the source position.
The typical positional uncertainty of Suzaku is 19$''$ \citep{Uchiyama2008},
while the systematic error of the peak determination is 8$''$.

The ASCA intensity map and the positions of XMM-Newton sources (3XMM DR4 version, \cite{Watson2009}) are 
also shown in figure 1. All the sources have possible X-ray counterparts that were found in the previous observations. 
We found possible radio and optical counterparts for Src2 and Src3 (SIMBAD database), 
but no counterpart in the third Fermi Large Area Telescope source catalog \citep{Acero2015}.
Using the NRAO VLA Sky Survey (NVSS) data \citep{Condon1998}, 
we also searched for extended radio emissions, but found no possible emission.
The possible counterparts are listed in table 2.

\begin{figure*}
  \begin{center}
        \includegraphics[width=16cm]{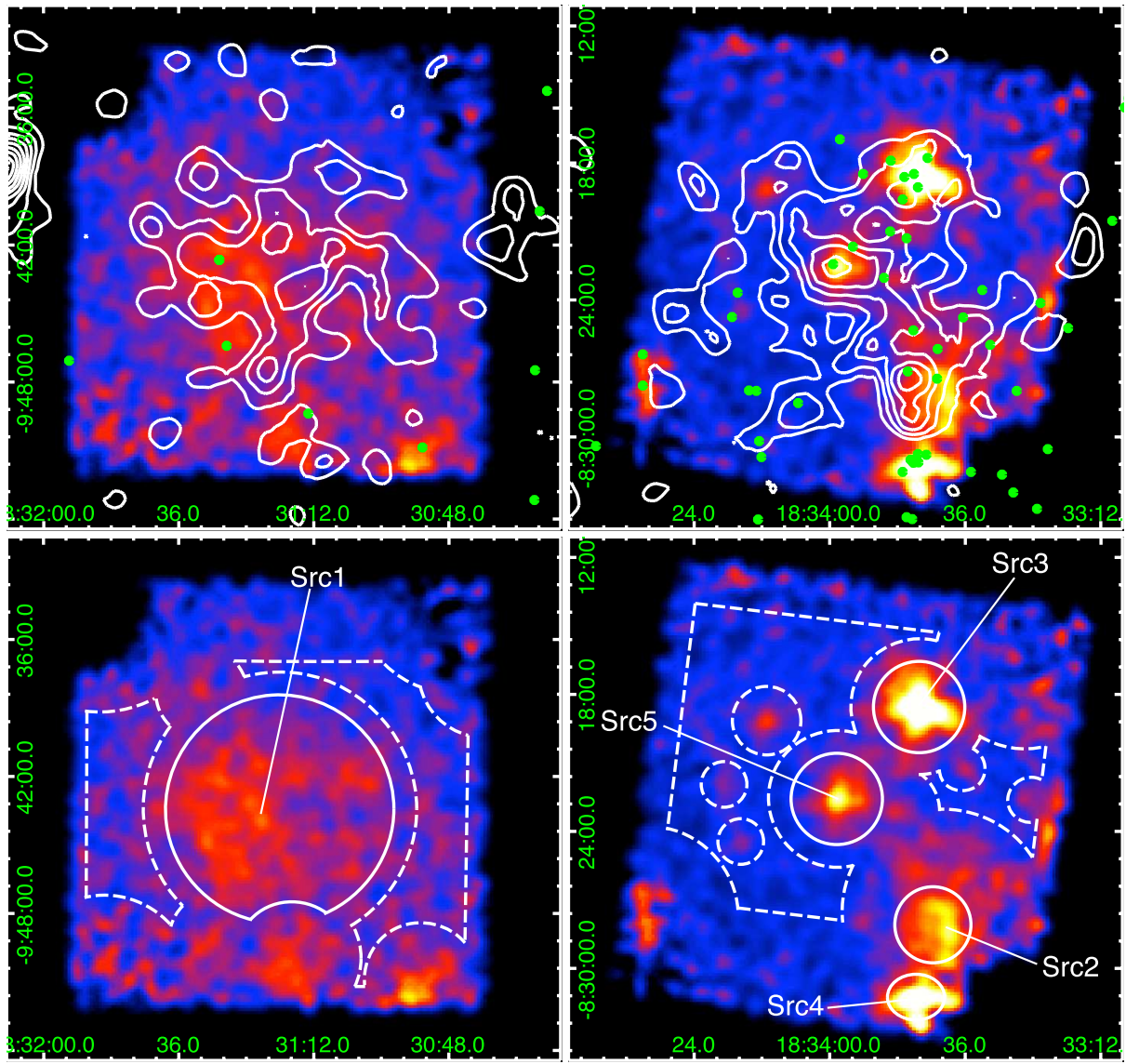}
  \end{center}
\caption{Upper: XIS image of G22.0$+$0.0 (left) and G23.5$+$0.1 (right) fields obtained 
in the 0.7--8 keV band (color). The coordinates are J2000.0.
The data of XIS 0, 1, and 3 were added. 
Background subtraction and vignetting correction are performed for the image. 
The intensity levels are linearly spaced.
The green dots show the positions of X-ray sources in the XMM-Newton Serendipitous Source Catalog 
(3XMM DR4 version, \cite{Watson2009}), while the white contours show the ASCA intensity map in the 0.7--8 keV band.
Lower: the same XIS images as the upper panel (color).
The white solid and dashed lines show source and background regions used for the spectral analysis, respectively.
}\label{fig:sample}
\end{figure*}

%
\begin{table*}[t]
\caption{Source positions and possible counterparts.}
\begin{center}
\begin{tabular}{lccclcc} \hline  
ID & Name &  Sky Position                   & Extent & Possible counterpart$^{\ast}$ & $\Delta\theta^{\dag}$ & References\\
Src             &  Suzaku J   &   (RA, Dec)$_{\rm J2000.0}$ &  &  & ($''$) &\\
\hline 
\multicolumn{7}{c}{G22.0$+$0.0 field} \\
\hline 
1 & 183121$-$0943&  (\timeform{18h31m21.s5}, \timeform{-9D43'57"}) & Yes &AX J183114$-$0943 (X) & 96 & 1 \\
    &                              &                                                                                          & &3XMM J183128.8$-$094239 (X) & 133 & 2 \\
    &                              &                                                                                         &  &3XMM J183127.5$-$094625 (X) & 173 & 2\\
\hline 
\multicolumn{7}{c}{G23.5$+$0.1 field} \\
\hline 
2& 183339$-$0828&  (\timeform{18h33m39.s1}, \timeform{-8D28'24"}) & Yes & CXO J183340.3$-$082830 (X) & 19 & 3\\ 
     &                                &                                                                                        & & = 2MASS J18334038$-$0828304 (IR) & 20 & 3\\
     &                                &                                                                                        & & 3XMM J183341.0$-$082727 (X) & 64 & 2\\
    &                                &                                                                                        & & = PSR B1830$-$08 (R)      & 56 & 3\\
     &                               &                                                                                         & &AX J183345$-$0828 (X) & 104 & 1\\
3 & 183344$-$0818&  (\timeform{18h33m44.s2}, \timeform{-8D18'32"}) & No & 3XMM J183345.1$-$081829 (X) & 14 & 2\\
   &                               &                                                                                         & & = SWIFT J183345.2$-$081831 (X) & 15 & 3 \\
     &                                &                                                                                        & & = BD-8 4632 (Opt) & 16 & 3\\
     &                                &                                                                                        & & = 2MASS J18334527$-$0818294 (IR) & 16 & 3\\
4& 183344$-$0831&(\timeform{18h33m44.s7}, \timeform{-8D31'15"}) & No & CXO J183344.4$-$083108 (X) & 9 & 3\\
     &                               &                                                                                       & & = 3XMM J183344.3$-$083107 (X) & 10 & 2\\
     &                               &                                                                                       & & = SGR J1833$-$0832 ($\gamma$) & 9 & 4\\
     &                               &                                                                                       & & 3XMM J183345.2$-$083108 (X) & 10 & 2\\
     &                               &                                                                                       & & 3XMM J183345.7$-$083100 (X) & 21 & 2\\
5 & 183358$-$0822&   (\timeform{18h33m58.s8}, \timeform{-8D22'34"}) & No & CXO J183359.4$-$082229 (X) & 11 & 3\\
   &                               &                                                                                         & & = 3XMM J183359.4$-$082226 (X) & 12 & 2\\
   &                               &                                                                                         & & = AX J183356$-$0822 (X) & 33 & 1\\
   \hline \\
\end{tabular}
\end{center}
\vspace{-10pt}
$^{\ast}$ X: X-ray, R: radio, IR: infrared, Opt: optical, and $\gamma$: $\gamma$-ray bands.\\
$^{\dag}$ Separation angle from the Suzaku position.\\
References: (1) \cite{Sugizaki2001}; (2) XMM-Newton Serendipitous Source Catalog (3XMM DR4 version); (3) Simbad database; 
(4) \cite{Gogus2010}.\\
\end{table*}

\subsection{Spectral analysis}

\begin{figure*}
  \begin{center}
        \includegraphics[width=8.0cm]{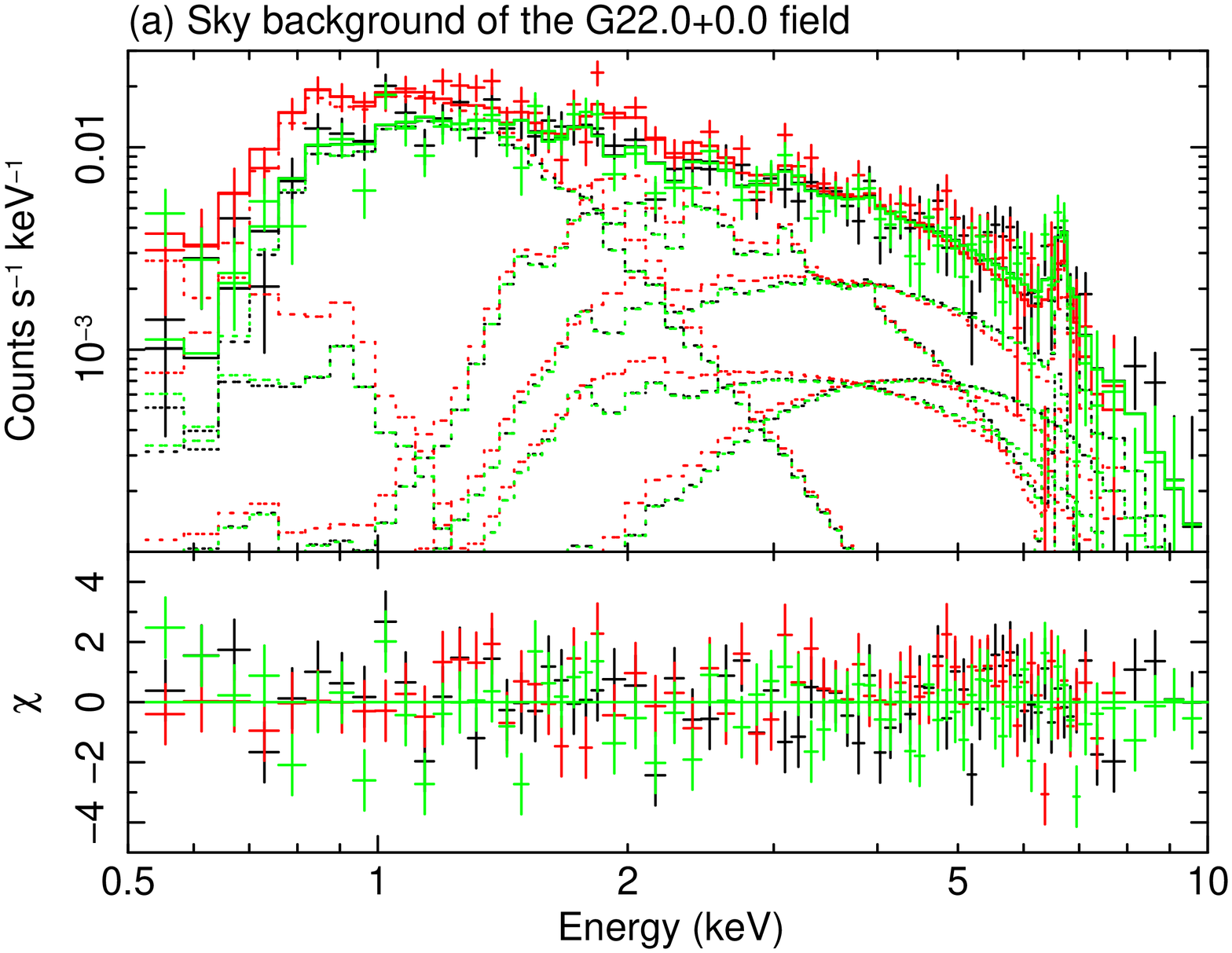}
        \includegraphics[width=8.0cm]{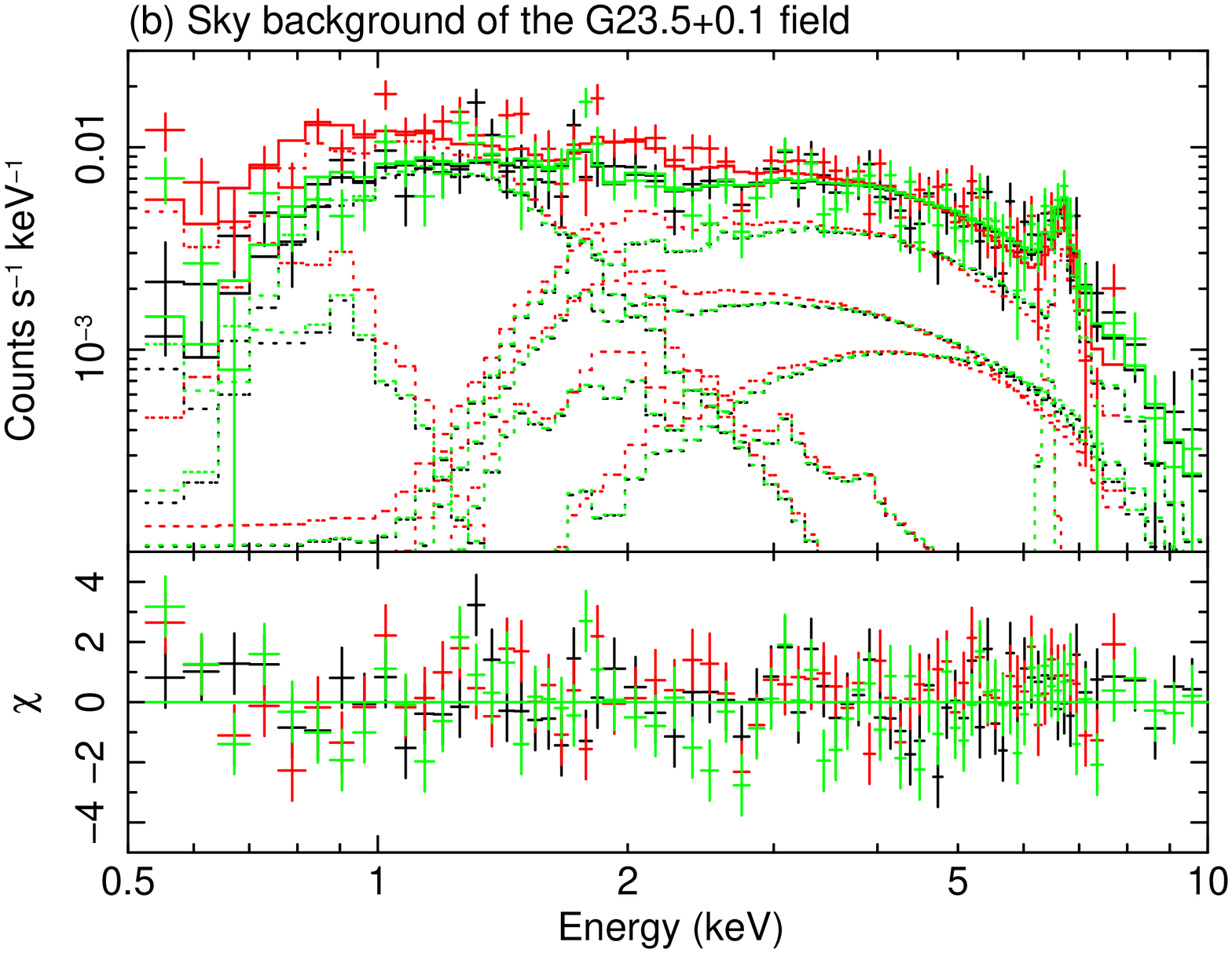}
  \end{center}
\caption{NXB-subtracted background spectra of the G22.0$+$0.0 (a) and G23.5$+$0.1 (b) fields 
and the residuals from the best-fit models (see table 3).
The XIS 0, 1, and 3 spectra were shown in black, red, and green, respectively.
The dotted lines show each component.

}\label{fig:sample}
\end{figure*}

Source spectra were extracted from the regions shown by the white solid lines in figure 1,
while sky background data were taken from the source free regions in the same FOV, 
shown by the dotted lines.
The source and the background regions were set to be common to XIS 0, 1, and 3.
The region near the calibration source at the FOV corners and the damaged part of XIS 0 
(north and south parts for G22.0$+$0.0 and G23.5$+$0.1 fields, respectively) were excluded.
Response files, Redistribution Matrix Files (RMFs)
and Ancillary Response Files (ARFs), were made using
{\tt xisrmfgen} and {\tt xissimarfgen}, respectively.
The non-X-ray background (NXB) for the source and the background spectra were extracted 
from the night-earth data (version 2014-06-01) using {\tt xisnxbgen} \citep{Tawa2008}.
The NXB was subtracted from the source and the background data.

\subsubsection{Sky background estimation}

%
\begin{table}[t]
\caption{The best-fit parameters of spectral analysis for the sky background.}
\begin{center}
\begin{tabular}{lcc} \hline  
Parameter & \multicolumn{2}{c}{Value}\\
\hline 
 & G22.0$+$0.0 field & G23.5$+$0.1 field\\
\hline 
$N_{\rm H, 1}$ (cm$^{-2}$) &  5.6$\times10^{21}$ (fixed)  & 5.6$\times10^{21}$ (fixed) \\
$kT_{\rm e, 1}$ (keV) &0.09 (fixed) & 0.09 (fixed) \\
Norm$^{\ast}$ (TP$_1$)& (3.5$\pm$2.0)$\times10^{-3}$ &(5.0$\pm$1.8)$\times10^{-3}$  \\
$kT_{\rm e, 2}$ (keV) & 0.59 (fixed) & 0.59 (fixed) \\
Norm$^{\ast}$  (TP$_2$)& (4.6$\pm$0.3)$\times10^{-5}$ & (2.2$\pm$0.2)$\times10^{-5}$ \\
$N_{\rm H, 2}$ (cm$^{-2}$) & (3.8$^{+0.4}_{-0.5}$)$\times10^{22}$  & (3.2$^{+0.6}_{-0.4}$)$\times10^{22}$ \\
$kT_{\rm e, L}$ (keV) &1.33 (fixed) & 1.33 (fixed) \\
Norm$^{\ast}$  (TP$_{\rm L}$)& (3.3$\pm$0.9)$\times10^{-5}$ & $<$9.3$\times10^{-6}$ \\
$kT_{\rm e, H}$ (keV) & 6.64 (fixed) & 6.64 (fixed) \\
Norm$^{\ast}$  (TP$_{\rm H}$)& (9.1$^{+2.6}_{-2.7}$)$\times10^{-6}$ &(1.2$^{+0.3}_{-0.2}$)$\times10^{-5}$ \\
Z$_{\rm Ar}^{\dag}$ (solar) &1.07 (fixed) & 1.07 (fixed) \\
Z$_{\rm others}^{\dag}$ (solar) & 0.81 (fixed) & 0.81 (fixed) \\
$\Gamma_{\rm RC}$ & 2.13 (fixed) & 2.13 (fixed) \\
$I_{\rm 6.4 keV}^{\ddag}$ & (1.4$\pm$1.3)$\times$10$^{-8}$& (2.4$^{+1.3}_{-1.1}$)$\times$10$^{-8}$\\
$N_{\rm H, 3}$ (cm$^{-2}$) & =2$\times$$N_{\rm H, 2}$ & =2$\times$$N_{\rm H, 2}$\\
$\Gamma_{\rm CXB}$ & 1.412 (fixed) & 1.412 (fixed) \\
Norm$_{\rm CXB}^{\S}$ & 8.2$\times$10$^{-7}$ (fixed) & 8.2$\times$10$^{-7}$ (fixed) \\

$\chi^2$/d.o.f. &  259.4/212 & 276.2/212   \\
\hline \\
\end{tabular}
\end{center}
\vspace{-12pt}
$^{\ast}$ Defined as 
10$^{-14}$$\times$$\int n_{\rm H} n_{\rm e} dV$ / (4$\pi D^2$),
where $n_{\rm H}$ is the hydrogen density (cm$^{-3}$), 
$n_{\rm e}$ is the electron density (cm$^{-3}$), and $D$ is the distance (cm). The unit is cm$^{-5}$ arcmin$^{-2}$.\\
$^{\dag}$ Relative to the solar value \citep{Anders1989}.\\
$^{\ddag}$ The unit is photons s$^{-1}$ cm$^{-2}$ arcmin$^{-2}$.\\
$^{\S}$ The unit is photons s$^{-1}$ cm$^{-2}$ keV$^{-1}$ arcmin$^{-2}$ at 1 keV.\\
\end{table}

Src1--5 are sources located on the Galactic plane, where a strong X-ray emission are found (e.g., \cite{Koyama1986}).
In order to minimize statistical uncertainty, we first evaluated sky background spectra and then modeled the sky background spectra 
with simulation.

The NXB-subtracted background spectra of the G22.0$+$0.0 and G23.5$+$0.1 fields are shown in figure 2.
\citet{Uchiyama2013} reported that the blank sky spectra on the Galactic plane were well represented by a
several component model as follows:

\medskip
[TP$_1$ ($kT_{\rm e}$=0.09keV)$+$TP$_2$ ($kT_{\rm e}$=0.59keV)]$\times$ABS1$+$

[TP$_{\rm L}$ ($kT_{\rm e}$=1.33keV)$+$TP$_{\rm H}$ ($kT_{\rm e}$=6.64keV)$+$RC]$\times$ABS2$+$

CXB$\times$ABS3

\medskip
\noindent
where TP, ABS, RC, and CXB show a thermal plasma model ({\tt vapec} in XSPEC), photoelectric absorption 
({\tt phabs} in XSPEC, \cite{bcmc1992}), a reflected component, and the Cosmic X-ray background.
RC is composed of a PL function and K$\alpha$ (6.4 keV) and K$\beta$ (7.05 keV) 
lines from neutral Fe. 
The K$\beta$ line intensity was fixed to be 0.125 $\times$ K$\alpha$ line intensity \citep{Kaastra1993}
and the equivalent width of K$\alpha$ line 
was fixed to be 457 eV \citep{Uchiyama2013}.
The metal abundances, temperatures, and the $N_{\rm H}$ value of ABS1
were fixed to those in \citet{Uchiyama2013} and 
the spectral parameters of the CXB were fixed to the values in \citet{Kushino2002}. 
The $N_{\rm H}$ value of ABS3 is assumed to be twice of that of ABS2.
Free parameters were normalizations of the TP models, the iron line intensity, and the $N_{\rm H}$ value of ABS2.
The abundance tables were taken from \citet{Anders1989}, 
while the line and continuum data of the thin thermal plasma were taken from the ATOMDB v3.0.1.
The XIS 0, 1, and 3 spectra were simultaneously fitted with the model.
The model represented the spectra 
with reduced $\chi^2$ ($\chi^2$/d.o.f.) values of 259.4/212=1.22 and 276.2/212=1.30 
for the G22.0$+$0.0 and G23.5$+$0.1 fields, respectively.
The best-fit model is plotted in figure 2, while the spectral parameters are listed in table 3.
The surface brightness is consistent with those in \citet{Uchiyama2013}.
Using the best-fit parameters and assuming a  10$^4$ times longer exposure time, 
we simulated the sky background spectra for each source region
and subtracted them from the NXB-subtracted source spectra.

\subsubsection{G22.0$+$0.0 field}

X-ray spectra of Src1 were extracted from a circle with a radius of 5$'$.
The background-subtracted spectra were displayed in figures 3.
The spectra exhibit no significant emission line from S, and Fe and 
the TP model fit gave an extremely high temperature ($kT\sim$60 keV).
Thus, the thermal model is unlikely. 
The PL model fit gave
a photon index and an $N_{\rm H}$ value of 0.95 and $<$3$\times$10$^{20}$ cm$^{-2}$, respectively, 
which are well consistent with the ASCA results \citep{Ueno2005}.
However, the reduced $\chi^2$ was 152.4/82=1.86 and positive residuals were clearly seen below 1 keV.  
In addition, we see an weak line-like feature at 1.2--1.3 keV, which would be a K-line from Mg.
Thus, we added a TP model ({\tt apec} model in XSPEC) modified by low-energy absorption.
The $N_{\rm H}$ values were set to be linked and the fit still gave a large reduced $\chi^2$ value of
111.7/80=1.40. 
Next, the $N_{\rm H}$ values were set to be unlinked and then
the fit was significantly improved ($\chi^2$/d.o.f.=93.1/79=1.18).
The best-fit parameters are listed in table 4, while the best-fit model is plotted in figure 3.

%
\begin{table*}[t]
\caption{The best-fit parameters of spectral analysis for Src1 in the G22.0$+$0.0 field.}
\begin{center}
\begin{tabular}{lccc} \hline  
Parameter & \multicolumn{3}{c}{Value}\\
\hline 
Model & PL$\times$ABS & (PL$+$TP)$\times$ABS & PL$\times$ABS$+$TP$\times$ABS\\
\hline 
$N_{\rm H}$ for PL  ($\times10^{22}$ cm$^{-2}$) &  
$<$0.03 & 0.8$^{+0.4}_{-0.2}$ & 2.6$^{+1.0}_{-0.9}$   \\
$\Gamma$   & 0.95$^{+0.08}_{-0.09}$ & 1.0$^{+0.2}_{-0.1}$ & 1.7$\pm$0.3   \\
$N_{\rm H}$ for TP ($\times10^{22}$ cm$^{-2}$) & 
--- & 0.8 (linked to $N_{\rm H}$ for PL) & 0.79$^{+0.15}_{-0.13}$  \\
$kT_{\rm e}$ (keV) & 
--- & 0.28$^{+0.09}_{-0.18}$ & 0.34$^{+0.11}_{-0.08}$  \\
Abundance$^{\ast}$ & --- & 1.0 (fixed) & 1.0 (fixed)   \\
$\chi^2$/d.o.f. &  152.4/82 & 111.7/80 & 93.1/79  \\
\hline \\
\end{tabular}
\end{center}
\vspace{-12pt}
$^{\ast}$ Relative to the solar value \citep{Anders1989}.\\
\end{table*}

\begin{figure}
  \begin{center}
       \includegraphics[width=8.0cm]{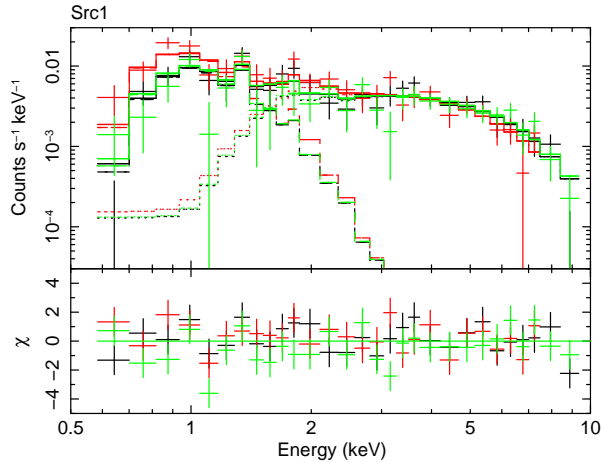}
  \end{center}
\caption{XIS spectra of Src1 and the residuals from the best-fit PL$\times$ABS$+$TP$\times$ABS model (see table 4).
The XIS 0, 1, and 3 spectra were shown in black, red, and green, respectively.
The dashed and the dotted lines show the TP and the PL models, respectively.
}\label{fig:sample}
\end{figure}

\subsubsection{G23.5$+$0.1 field}

X-ray spectra of Src2, Src3, and Src5 were extracted from a circle with a radius of 100$''$, 2$'$, and 2$'$, respectively, while 
those of Src4 were extracted from an ellipse with a size of \timeform{2.'5}$\times$2$'$.
Since Src4 is located close to the calibration source of FIs and the damaged area of XIS 0,
only the XIS 1 data were utilized.
The background-subtracted spectra were shown in figure 4.
The spectra were fitted with either the TP or the PL model modified by low-energy absorption. 
The best-fit parameters are listed in table 5 and the best-fit models are plotted in figure 4.

Applying the absorbed PL model for the Src5 spectra,
we clearly found positive residuals at 6--7 keV, and hence
we added an emission line with a line width of null.
The $\Delta {\chi}^2$ value was 7.5, showing that 
the additional emission line model is at a 97 \% confidence level.
The center energy was determined to be 6.80$^{+0.12}_{-0.07}$ keV, 
which indicates that the line is K-shell transition line from highly ionized iron.
Thus, the emission from Src5 is thin thermal.
An equivalent width (EW) was estimated to be 320$\pm$200 eV.

%
\begin{table*}[t]
\caption{The best-fit parameters of spectral analysis for Src2--5 in the G23.5$+$0.1 field.}
\begin{center}
\begin{tabular}{lcccccccc} \hline  
Parameter & \multicolumn{7}{c}{Value}\\
                    & \multicolumn{2}{c}{Src2} & \multicolumn{2}{c}{Src3} & \multicolumn{2}{c}{Src4} & \multicolumn{2}{c}{Src5} \\
\hline 
Model & PL$\times$ABS & TP$\times$ABS & PL$\times$ABS & TP$\times$ABS & PL$\times$ABS & TP$\times$ABS & PL$\times$ABS & TP$\times$ABS \\
\hline
$N_{\rm H}$  ($\times10^{22}$ cm$^{-2}$) & 
3.5$^{+1.1}_{-0.8}$ & 2.9$^{+0.7}_{-0.6}$ & 0.45$^{+0.08}_{-0.09}$ & 0.09$^{+0.06}_{-0.04}$ & 
21$^{+9}_{-7}$& 18$^{+7}_{-5}$ & 2.5$^{+1.9}_{-1.2}$ & 5.1$^{+1.4}_{-1.2}$ \\
$\Gamma$ / $kT_{\rm e}$ (keV)   & 
2.4$^{+0.5}_{-0.4}$ & 4.4$^{+2.7}_{-1.3}$ & 5.2$^{+0.6}_{-0.5}$ & 0.81$^{+0.15}_{-0.05}$ & 
4.8$^{+1.9}_{-1.4}$ &1.6$^{+1.1}_{-0.6}$ &0.9$\pm$0.4 & $>$10 \\
Abundance$^{\ast}$ &
 ---  & $<$0.24 & --- & 0.07$\pm$0.02 & --- & $<$0.58 & --- & 1.0 (fixed) \\
$\chi^2$/d.o.f. & 78.3/62 &76.3/61 &  242.4/122 &138.5/121 & 18.2/28 & 17.9/27 & 53.2/49 & 51.5/49\\
\hline \\
\end{tabular}
\end{center}
\vspace{-12pt}
$^{\ast}$ Relative to the solar value \citep{Anders1989}.\\
\end{table*}

\begin{figure*}
  \begin{center}
       \includegraphics[width=8.0cm]{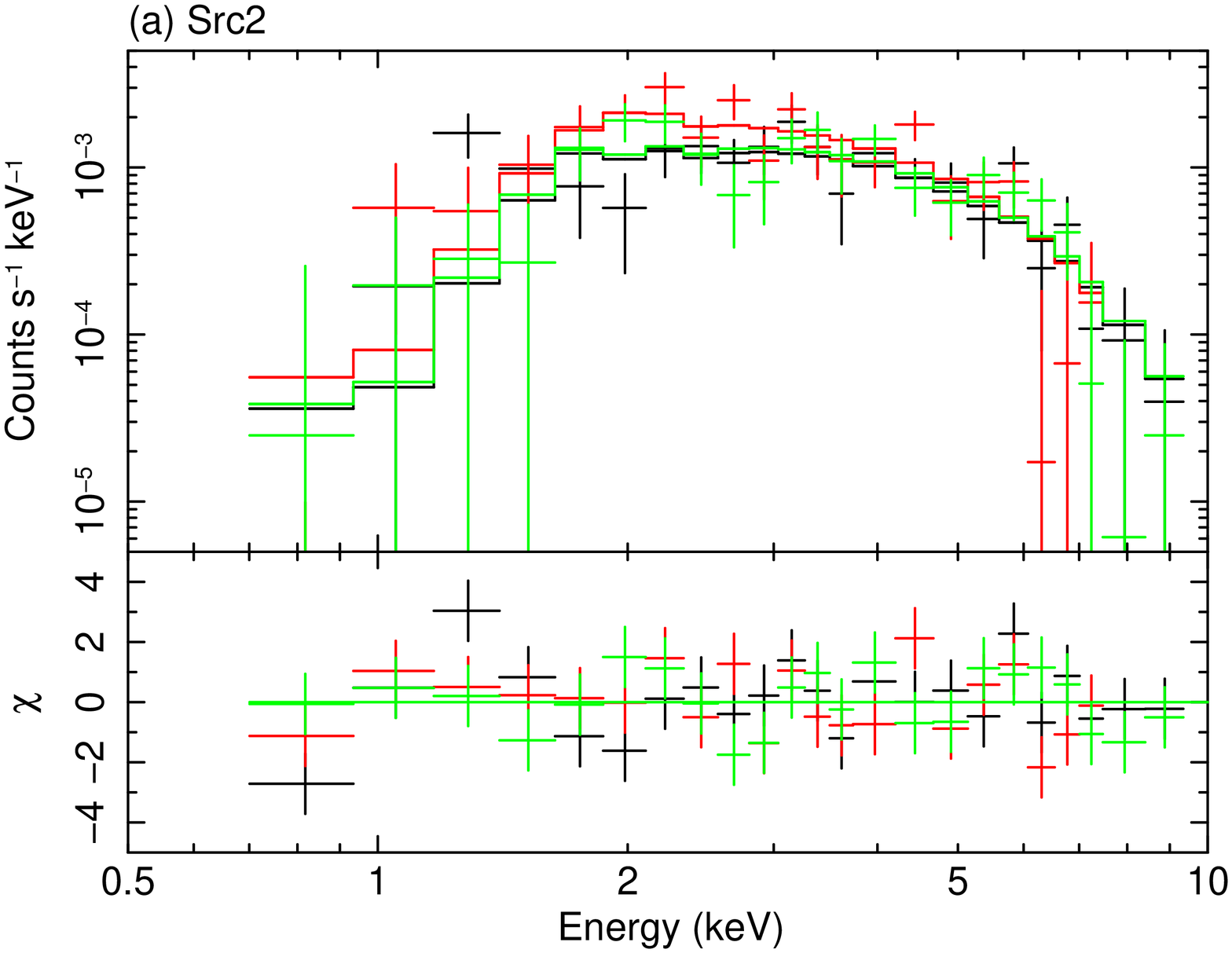}
       \includegraphics[width=8.0cm]{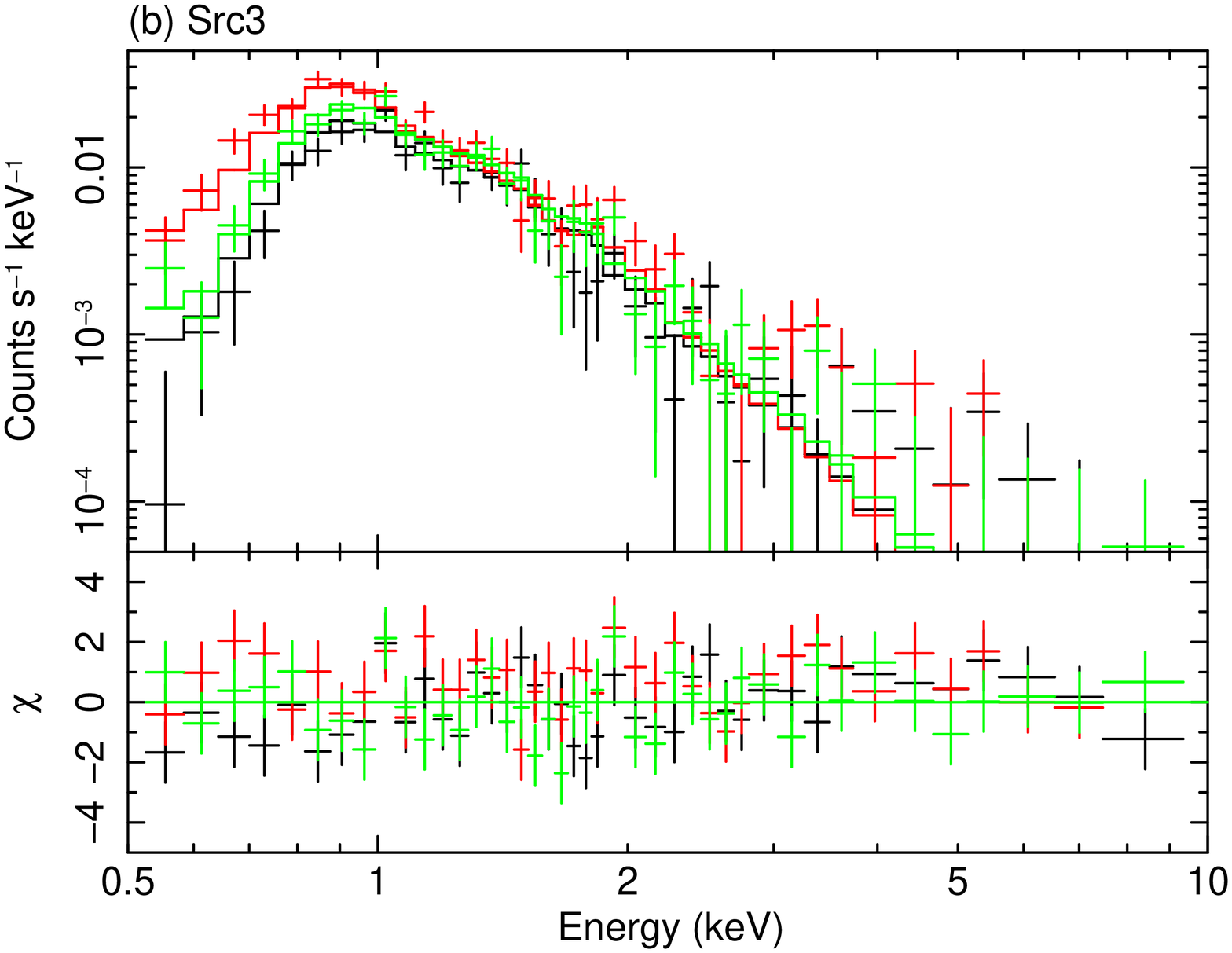}
       \includegraphics[width=8.0cm]{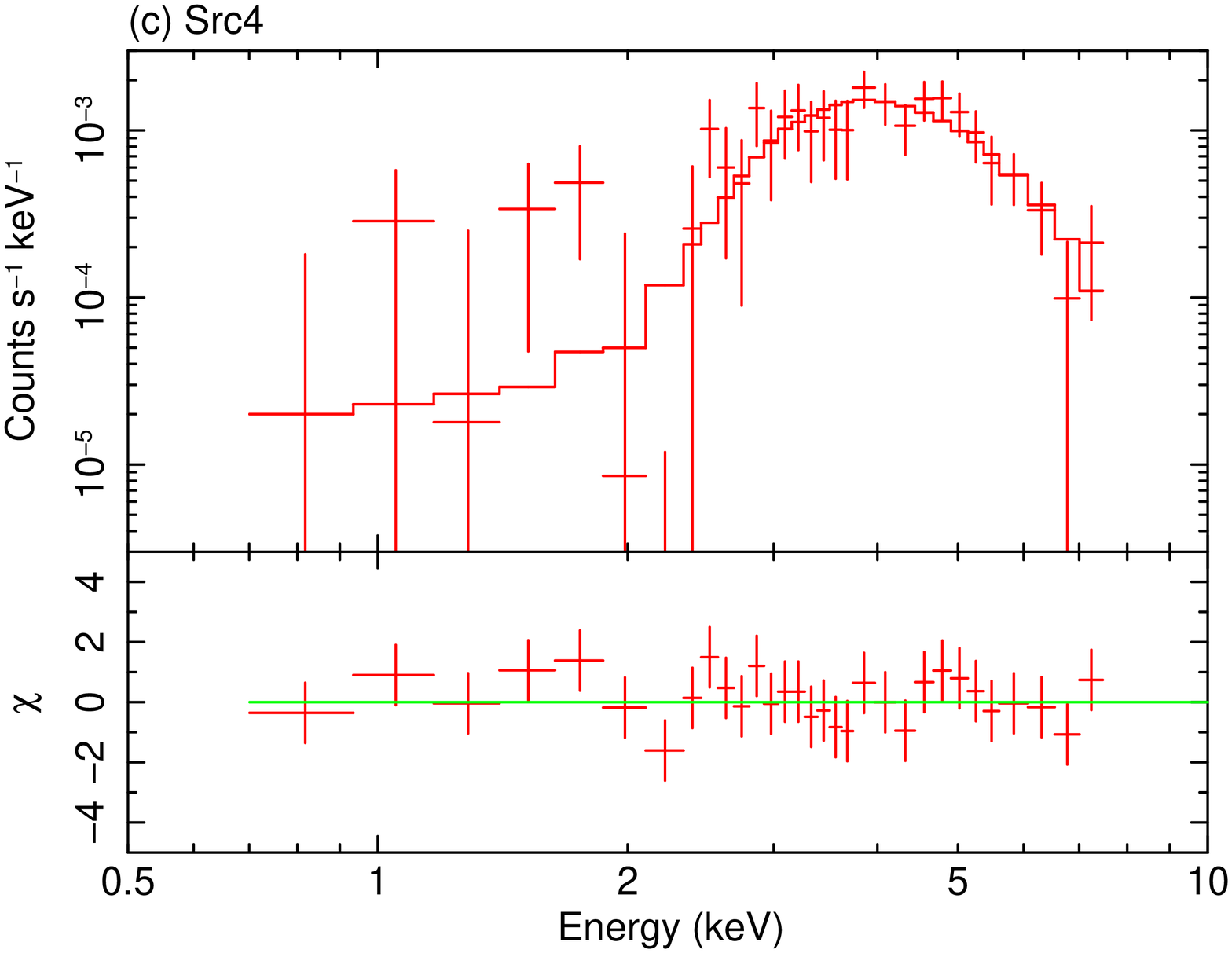} 
       \includegraphics[width=8.0cm]{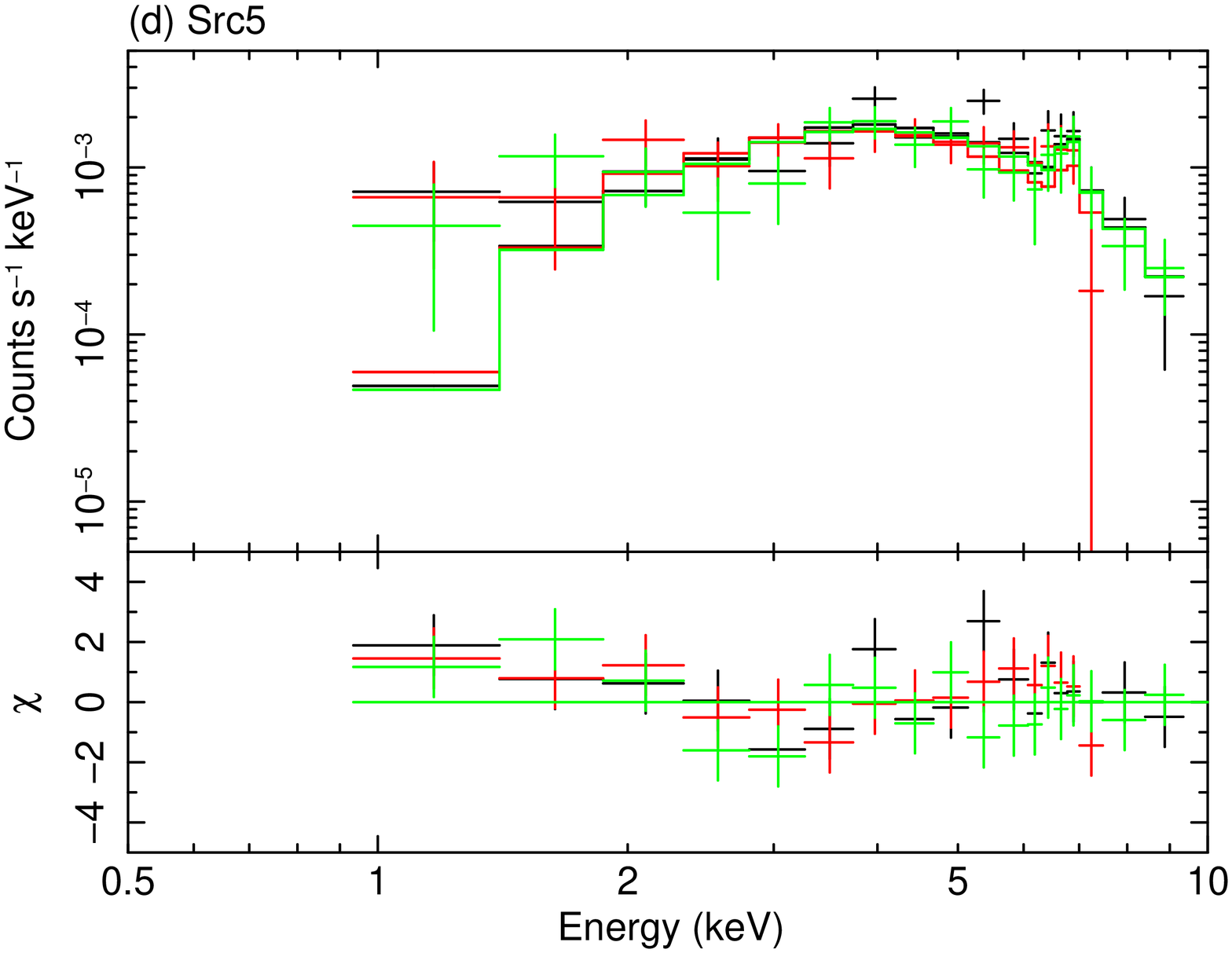}
  \end{center}
\caption{XIS spectra of Src2 (a), Src3 (b), Src4 (c), and Src5 (d) and the residuals from the best-fit models:
the PL model for Src2 and Src4 and the TP model for Src3 and Src5 (see table 5).
The XIS 0, 1, and 3 spectra were shown in black, red, and green, respectively.
}\label{fig:sample}
\end{figure*}

\section{Discussion}

\subsection{Extended sources}

\subsubsection{Src1}

The spatial structure and the derived spectral parameters are very
similar to the ASCA results \citep{Ueno2005,Ueno2006}. 
Thus, Src1 is identified with AX J183114$-$0943.
The spectra of Src1 are represented by two component model, the PL and TP model.
The TP model is dominant below 1 keV (see figure 3), 
but no apparent source was found in the 0.7--1 keV band image.
The $N_{\rm H}$ value of the TP model is about one-third of  that of the PL model.
These indicate
that the thermal component is unrelated with the PL component and 
is a foreground emission located in the same line-of-sight. 
Here, we focus on the PL component.

The source region contains two XMM-Newton sources.
The fluxes of 3XMM J183128.8$-$094239 and 3XMM J183127.5$-$094625 
in the 0.2--12.0 keV energy band are 5.0$\times$10$^{-14}$ erg s$^{-1}$ cm$^{-2}$ and 
9.5$\times$10$^{-14}$ erg s$^{-1}$ cm$^{-2}$, respectively (3XMM DR4 version).
The flux of the PL component is estimated to be $\sim$10$^{-12}$ erg s$^{-1}$ cm$^{-2}$ 
in the same energy band.
Thus, we conclude that the contribution of XMM-Newton sources is small and 
most of X-rays ($\sim$85\%) originate from the extended emission.

The Galactic HI column density ($N_{\rm HI}$) along the line-of-sight to Src1
is $N_{\rm HI}$=2.0$\times$10$^{22}$ cm$^{-2}$ \citep{Dickey1990} or 
$N_{\rm HI}$=1.6$\times$10$^{22}$ cm$^{-2}$ \citep{Kalberla2005}.
The CO intensity at the source position (160--210 K km s$^{-1}$; \cite{Dame2001}) and the 
conversion factor to the $N_{\rm H_2}$ value [(1.8$\pm$0.3)$\times$10$^{20}$ cm$^{-2}$ K$^{-1}$ km$^{-1}$ s; \cite{Dame2001}]
lead to the Galactic H$_2$ column density ($N_{\rm H_2}$) of (3.3$\pm$0.7)$\times$10$^{22}$ cm$^{-2}$. 
Thus, the total $N_{\rm H}$, $N_{\rm H}$=$N_{\rm HI}+2N_{\rm H_2}$, is estimated to be 
(8.4$\pm$1.4)$\times$10$^{22}$ cm$^{-2}$. 
The $N_{\rm H}$ value of Src1 is about 30 \%
of the total $N_{\rm H}$ through the Galaxy. 
If the line-of-sight length of the $N_{\rm HI}$ and $N_{\rm H_2}$ estimation in
the radio band is assumed to be 17 kpc, double the distance
from the Sun to the Galactic center, the distance of Src1 is 5.2$\pm$2.2 kpc.

The photon index of 1.7$\pm$0.3 is typical values of PWNe
($\Gamma\sim$2, e.g., \cite{Possenti2002,Kargaltsev2008}) rather than those of SN1006-like SNRs 
($\Gamma\sim$2.5--3, e.g., \cite{Koyama1995, Bamba2005}). 
The X-ray image exhibits not shell-like but center-filled morphology.
Thus, Src1 is likely to be a PWN.
Taking account of the uncertainty of the distance, 
we estimated the X-ray luminosity in the 2--10 keV band and the angular size 
to be (0.9--6.6)$\times$10$^{33}$ erg s$^{-1}$ and 9--22 pc, respectively.
\citet{Possenti2002} and \citet{Mattana2009} presented
the empirical relation between X-ray luminosity and the characteristic age of pulsars in PWNe.
\citet{Bamba2010} showed the relation between the characteristic age and the nebula size in the X-ray band.
Comparing with the relations, we found that 
the estimated size and luminosity are consistent with those of PWNe 
with a larger characteristic age of $\sim$10$^4$--10$^5$ yr.

Thus, all the facts support the idea that Src1 is a new candidate of an old PWN. 
However, we found no apparent high-energy compact source.
To establish the scenario, search for the high-energy source is essential.

\subsubsection{Src2}

The source region contains a point source CXO J183340.3$-$082830=2MASS J18334038$-$0828304. 
The flux was estimated to be $\sim$7$\times$10$^{-14}$ erg s$^{-1}$ cm$^{-2}$ in the 0.3--8 keV band 
(Chandra XAssist Source List, Ptak \& Griffiths 2003),
which is $\sim$16\% of the observed flux of Src2.
The discrepancies of the flux and spatial structure indicate that CXO J183340.3$-$082830 is not a main counterpart of Src2.
On the other hand, the spectral parameters of the PL model fit are fully consistent with those of source 8 in \citet{Kargaltsev2012}, 
and are in agreement with those in \citet{Esposito2011} within the errors. 
Furthermore,  the size is also consistent.
Thus, these sources are identical.

Using $N_{\rm HI}$=2.1$\times$10$^{22}$ cm$^{-2}$ \citep{Dickey1990} or 
$N_{\rm HI}$=1.7$\times$10$^{22}$ cm$^{-2}$ \citep{Kalberla2005}, 
the CO intensity at the source position of 180--230 K km s$^{-1}$ \citep{Dame2001}, 
we can estimate $N_{\rm H}$=(9.4$\pm$1.6)$\times$10$^{22}$ cm$^{-2}$ along the line-of-sight to Src2. 
The $N_{\rm H}$ value of (3.5$^{+1.1}_{-0.8}$)$\times$10$^{22}$ cm$^{-2}$ of Src2 leads to the distance of 6.4$\pm$2.3 kpc.

Src2 is a Galactic source with an extension of $\sim$3$'$--4$'$, suggesting an SNR or a PWN.
If the X-rays have thermal origin,
the higher temperature derived from the TP model fit suggests a young SNR such as Tycho and Kepler SNRs
(e.g., \cite{Hwang1997,Kinugasa1999}). 
However, we found no Si, S, and Fe emission lines that are typically seen in the SNR spectra;
the abundance is very small (less than 0.24 solar, 90\% confidence level). 
The photon index of 2.4$^{+0.5}_{-0.4}$ is a typical value of SN 1006-like SNRs or PWNe 
\citep{Possenti2002,Kargaltsev2008,Koyama1995, Bamba2005}. 
The X-ray luminosity in the 2--10 keV band and the size are estimated to be 
(1.2--6.5)$\times$10$^{33}$ erg s$^{-1}$ and 4--9 pc, respectively.
The angular size is small for a non-thermal SNR with the luminosity level \citep{Nakamura2012}.
On the other hand, 
the observed properties of Src2 are well consistent with those of PWNe with the characteristic age of $\sim$10$^4$ yr
\citep{Possenti2002,Mattana2009,Bamba2010}. 
Thus, Src2 is likely to be an old PWN.

\citet{Esposito2011} and \citet{Kargaltsev2012} proposed that 
the extended nebula is a PWN generated by B1830$-$08=PSR J1833$-$0827.
The estimated distance of Src2 is consistent with that of B1830$-$08, and hence our results support the scenario. 
Considering the characteristic age of B1830$-$08 ($\tau$=147 kyr), the size is a little bit small \citep{Bamba2010}.
Some mechanisms to prevent the nebula from expanding, such as interaction with SNR reverse shock \citep{Gaensler2006},
may work or there may be 
a low surface brightness nebula that we have not found.

\subsection{Point sources}

\subsubsection{Src3}

Src3 exhibits softer spectra than the other sources.
The TP model fit is acceptable (see table 5) and 
the spectral parameters are essentially the same as those in \citet{Kargaltsev2012}.

Using the same method as those for Src1 and Src2, 
we can estimate the distance of Src3 to be 170$\pm$100 pc, 
and hence the luminosity is calculated to be (0.3--4.0)$\times$10$^{30}$ erg s$^{-1}$ in the 0.5--10 keV band.
The spectral parameters and the luminosity are well consistent with those of late-type stars (e.g., \cite{Schmitt1990}). 
Based on the SIMBAD database, 
$B$ and $V$ band magnitudes of BD-8 4632 (the optical counterpart of Src3) are 10.66 and 9.97, respectively.
The observed $N_{\rm H}$ value and the empirical 
relation between $N_{\rm H}$ value and the color excess, $E_{\rm B-V}$, of 
$N_{\rm H}$/$E_{\rm B-V}$=(6.8$\pm$1.6)$\times$10$^{21}$ cm$^{-2}$ mag$^{-1}$ \citep{Ryter1975} 
give the color, $B-V$, to be 0.55$\pm$0.09. 
These facts show that Src3 is a F--G type star.
This is the same conclusion as that in \citet{Kargaltsev2012}.

\subsubsection{Src4}

Src4 is most likely to be SGR J1833$-$0832 discovered on 2010 March 19 
because the best-fit spectral parameters are in agreement with those of \citet{Gogus2010} within the errors. 
The Suzaku observation was carried out on $\sim$220 days after the outburst.
The spectra of SGR J1833$-$0832 are well represented by an absorbed blackbody model \citep{Gogus2010,Esposito2011}.
Thus, we also applied the model and obtained 
the temperature of 0.83$^{+0.23}_{-0.17}$ keV, the absorption column density of (1.4$^{+0.7}_{-0.5}$)$\times$10$^{23}$ cm$^{-2}$, 
and the observed flux in the 2--10 keV of 6$\times$10$^{-13}$ erg s$^{-1}$ cm$^{-2}$ ($\chi^2$/d.o.f.=17.7/28).
Compared with the results in \citet{Gogus2010} and \citet{Esposito2011}, 
the temperature and the flux are gradually decreased after the Swift and XMM-Newton observations.

\subsubsection{Src5}

Src5 is identified with source 3 (=3XMM J183359.4$-$082226) in \citet{Kargaltsev2012}. 
\citet{Kargaltsev2012} proposed that the source is either 
a neutron star binary (NSB), an active galactic nucleus (AGN), or a cataclysmic variable (CV).
However, no spectral information on 3XMM J183359.4$-$082226 has been reported so far.
The Suzaku observation revealed that the spectra 
are well represented by the TP model with a high temperature of $>$10 keV and an weak Fe-K line at 6.8 keV
($EW$=320$\pm$200 eV).
These features are not found in NSB and AGN (e.g., \cite{White1995,Turner1989}),
but are well consistent with those of CVs (e.g., \cite{Ezuka1999}).
The observed $N_{\rm H}$ value suggests the distance to be 7.9$\pm$2.5 kpc. 
Then the luminosity in the 2--10 keV band is estimated to be (1.9--6.9)$\times$10$^{33}$ erg s$^{-1}$, 
which is also in the range of CVs.
Thus, all the facts indicate that Src5 is most likely to be a CV.
 
\section*{Acknowledgement}

The authors are grateful to all members of the Suzaku team. 
This work was supported by JSPS KAKENHI Grant Number 24540232.
This research has made use of the SIMBAD database,
operated at CDS, Strasbourg, France.


\end{document}